\documentclass[review]{elsarticle}

\usepackage{psfrag,graphicx}
\usepackage{dcolumn}
\usepackage{amsmath,amssymb}
\usepackage{bm}
\usepackage{amsfonts,amssymb,amsmath}        

\newcommand{\be}{\begin{equation}}
\newcommand{\ee}{\end{equation}}
\newcommand{\bq}{\begin{eqnarray}}
\newcommand{\eq}{\end{eqnarray}}

\newcommand{\ket}[1]{\left | \, #1 \right\rangle}
\newcommand{\bra}[1]{\left \langle #1 \, \right |}
\newcommand{\rf}[1]{(\ref{#1})}
\begin{document}

\title{Engineering complex topological memories from simple Abelian models}

\author[lds]{James R. Wootton \corref{cor1}}
\ead{phyjrw@leeds.ac.uk}

\author[lds]{Ville Lahtinen}
\author[par]{Benoit Doucot}
\author[lds]{Jiannis K. Pachos}

\cortext[cor1]{Corresponding author}

\address[lds]{School of Physics and Astronomy, University of Leeds, Woodhouse Lane, Leeds LS2 9JT, UK}
\address[par]{Laboratoire de Physique Theorique et Hautes Energies, Universite Paris 6 et 7, Paris, France}

\date{\today}

\begin{abstract}

In three spatial dimensions, particles are limited to either bosonic or fermionic statistics. Two-dimensional systems, on the other hand, can support anyonic quasiparticles exhibiting richer statistical behaviours. An exciting proposal for quantum computation is to employ anyonic statistics to manipulate information. Since such statistical evolutions depend only on topological characteristics, the resulting computation is intrinsically resilient to errors. So-called non-Abelian anyons are most promising for quantum computation, but their physical realization may prove to be complex. Abelian anyons, however, are easier to understand theoretically and realize experimentally. Here we show that complex topological memories inspired by non-Abelian anyons can be engineered in Abelian models. We explicitly demonstrate the control procedures for the encoding and manipulation of quantum information in specific lattice models that can be implemented in the laboratory. This bridges the gap between requirements for anyonic quantum computation and the potential of state-of-the-art technology.

\end{abstract}

\begin{keyword}
Topological models, Anyons, Fault-tolerant quantum computation
\PACS {03.67.Lx, 03.67.Pp, 05.30.Pr, 73.43.Lp, 03.65.Vf}
\end{keyword}

\maketitle

\section{Introduction}
It has been theoretically demonstrated that quantum
computation can perform tasks that are virtually impossible with classical
computers. Physical realizations are presently hindered by environmental and
control errors that do not allow demonstration of quantum computation beyond
the classical limit \cite{DV1,DV2}. Schemes based on topologically ordered systems promise to drastically resolve this
problem by employing physical principles as well as algorithmic
constructions. Their aim is to encode and manipulate information in a way
that is intrinsically resilient to errors, thus allowing fault-tolerant quantum
computation.

Topological quantum computation employs anyons, the quasiparticles of topological models \cite{Wilczek,Anyons1,Anyons2,Double1}. Non-Abelian anyons have complex behavior well suited to encoding and processing quantum information, but Abelian anyons are simpler. Here we bridge the gap, using well established control techniques to engineer the so-called `non-Abelian like' encoding of \cite{tqc,thesis} using an Abelian lattice model. Utilizing certain symmetries of the Hamiltonian, which may be enforced by single spin interactions, we interpret certain states of Abelian anyons in terms of new quasiparticles \cite{Wootton}. These quasiparticles exhibit non-Abelian like behavior, though they do not not constitute a full mapping to non-Abelian anyons. Using the new quasiparticles, quantum information can be stored in the analogues of fusion spaces and manipulated using single spin measurements and adiabatic techniques. The addition of non-topological operations, namely measurements of single spins in the underlying lattice, leads to universal quantum computation \cite{Lots,Rauss,me,qpl,tqc}. State purification is employed to efficiently eliminate errors introduced by these manipulations. All employed operations, including measurements, always produce eigenstates of the Hamiltonian, allowing the energy gap to remain intact at all times and ensuring fault tolerance \cite{Brennen08}.

Though our method is general, we focus on the quantum double models \cite{Double1,Double2}, analytically tractable topological lattice models described by the stabilizer formalism \cite{Gottesman}. Specifically we consider the $D(Z_6)$ model, which an be realized with Josephson junctions \cite{Doucot}. An experiment is proposed to demonstrate the quantum memory of our scheme, as well as our method to enforce symmetries.

\section{Non-Abelian like memories}

Non-Abelian anyons have a multiplicity of possible outcomes when fused. The states governing these outcomes belong to the so-called fusion space of the anyons, which is completely inaccessible to any local or LOCC measurement or error. As such, this space provides the perfect place to store quantum information in a non-local, and yet fault-tolerant, manner. In \cite{tqc,thesis}, some of us studied such a non-Abelian model and its corresponding quantum memory. It was found that, in addition to the `true non-Abelian memory', an alternative method of encoding within the fusion space could be realized. This `non-Abelian like memory' shares many features with the true memory, including the use of the fusion space, but is not truly non-local. Instead it can be described as `delocalized' since it is inaccessible to local operations alone, giving a measure of protection, but can be accessed through LOCC, allowing a simpler realization.

Specifically the model considered in \cite{tqc,thesis} has anyons $\Phi$ and $\Lambda$ with the fusion rules $\Phi \times \Phi = 1+\Lambda+\Phi$ and $\Phi \times \Lambda = \Phi$. A qubit can therefore be stored in two $\Phi$ pairs. The qubit state $\ket{0}$ can be encoded in the state where both pairs fuse to the vacuum, and $\ket{1}$ can be stored in two pairs that fuse to a $\Lambda$. The logical $X$ operation is then implemented by fusing a $\Lambda$ with a $\Phi$ from each pair. The difference between the true and non-Abelian like memories come from the way this fusion is performed. In the non-Abelian like memory, fusion is performed by simply moving the $\Lambda$ onto the same site as a $\Phi$. This causes the $\Lambda$ to disappear, and so locally seems to perform the fusion. However, LOCC measurement of both the $\Phi$ and its partner can still be used to detect its presence, and hence measure the logical state. This memory therefore stores the information in a way that cannot be detected with local measurements alone, but can be detected with LOCC. For the true non-Abelian memory, in addition to moving the $\Lambda$ onto the same site as the $\Phi$, an entangling operation is applied to all spins surrounding the site. This then harnesses the non-Abelian structure of the underlying spin lattice to make the $\Lambda$ invisible even to LOCC measurements, and so truly fused. This then gives the non-local encoding expected from non-Abelian models. However, the need for the additional entangling operation makes the realization of this encoding more complex.

Here we build on this previous work by showing that corresponding non-Abelian like memories can also be implemented using Abelian anyons. In place of the $\Phi$ and $\Lambda$ anyons, we define quasiparticles $\phi$, $\overline\phi$ and $\lambda$ for which the non-Abelian like memories can be defined in a completely equivalent way. Furthermore, by the use of single spin interactions, we demonstrate how the encoding can be strengthened to give a fault-tolerance comparable to the true non-Abelian memory. A complementary presentation of this work can also be found in \cite{thesis}.

\section{The $D(Z_6)$ model}

The $D(Z_6)$ anyon model is defined on an oriented
two-dimensional square lattice \cite{Double1,Double2}. On each edge there resides a
six-level spin spanned by the states $\ket{g}_i$, where $g=0,..,5$ labels an
element of $Z_6$, the cyclic group of six elements. The generalized Pauli operators for these spins are,
\be
\sigma^x_i = \sum_{g \in Z_6} \ket{g+1}_i \bra{g}, \, \sigma^z_i = \sum_{g \in Z_g} e^{-i \pi g /3} \ket{g}_i \bra{g}.
\ee

We choose the orientations of our lattice to point upwards for each vertical link and right for each horizontal link. We can now define the following operators,
\bq \label{AB}
A (v) = \sigma^x{}^\dagger_j \sigma^x_k{}^\dagger \sigma^x_l \sigma^x_m, \,\,\, B (p) = \sigma^z{}^\dagger_j \sigma^z_k \sigma^z_l \sigma^z{}^\dagger_m.
\eq
The edges $j,k,l,m$ are those sharing the plaquette $p$ or vertex $v$. We take $j$ to be the edge to the top of the plaquette or vertex and the rest to proceed clockwise from this. The Hamiltonian of the model is given by,
\be \label{H}
H = -\Delta \left( \sum_v P_1 (v) + \sum_p P_1 (p) \right),
\ee
where $\Delta > 0$.  A state within the ground state space, $\ket{\textrm{gs}}$, is then defined by the following projectors,
\be \nonumber
P_1 (v) = \frac{1}{6} \sum_{g \in Z_6} \big(A (v) \big)^g, \,\,\, P_1 (p) = \frac{1}{6} \sum_{g \in Z_6} \big( B (p) \big)^g,
\ee
such that $P_1(v)\ket{\textrm{gs}}=P_1(p)\ket{\textrm{gs}}=\ket{\textrm{gs}}$ for all $v$ and $p$. All $A(v)$ and $B(p)$ operators mutually commute, and so may be described as stabilizers of a stabilizer code \cite{Gottesman}. The ground state space is then identified with the stabilizer space. The elementary excitations of $D(Z_6)$ are the anyons $e_g$ and $m_g$ for $g=1,..,5$, and correspond to violations of the vertex and plaquette stabilizers, respectively. The absence of an anyon is referred to as the vacuum, which is denoted $1$. The fusion rules of these anyons, which govern the result when two anyons are moved to the same vertex or plaquette, are,
\be \label{fuse}
e_g \times e_h = e_{g+h}, \,\,\, m_g \times m_h = m_{g+h}.
\ee
The addition is done mod $6$, with $e_0$ and $m_0$ identified with the vacuum, $1$.

The projectors for the anyon states are given by,
\bq \label{em} \nonumber
P_{e_g} (v) &=& \frac{1}{6} \sum_{h \in Z_6} e^{-i \pi g h /3} \big( A (v) \big)^h, \\
P_{m_g} (p) &=& \frac{1}{6} \sum_{h \in Z_6} e^{-i \pi g h /3} \big( B (p) \big)^h.
\eq
When the system is in the state $\ket{\psi}$, a quasiparticle of type $a$ at vertex $v$ (plaquette $p$) is defined by $P_a(v) \ket{\psi} = \ket{\psi}$ ($P_a(p) \ket{\psi} = \ket{\psi}$). An important feature of the Hamiltonian \rf{H} is that it assigns the same energy, $\Delta$, to all $e_g$ anyons, and also to all $m_g$ anyons. This symmetry is not a necessary for a valid $D(Z_6)$ Hamiltonian, but it is necessary for the scheme we propose. Later we present a simple method to protect this symmetry. 

To encode quantum information in a protected way, we define new types of quasiparticle. We focus on the vertex excitations, while the plaquette excitations are similarly defined. Let us introduce the projectors,
\bq \label{P} \nonumber
P_{\lambda} (v) &=& P_{e_3} (v), \quad P_{\phi} (v) = P_{e_1} (v)  + P_{e_4} (v) \\
P_{\overline{\phi}} (v) &=& P_{e_2} (v) + P_{e_5} (v).
\eq
In terms of the anyons of $D(Z_6)$, the quasiparticle $\phi$ is an $e_1$ or $e_4$ and its antiparticle $\overline \phi$ is an $e_2$ or $e_5$. The quasiparticle $\lambda$ is directly identified with $e_3$. The fact that the Hamiltonian assigns the same energy to each of these anyons allows arbitrary states of these quasiparticles to be eigenstates. Measuring the stabilizer code with the projectors in \rf{P} rather than \rf{em} extracts less information about the type of anyon present. This is equivalent to 'holes' in the code, where stabilizers are not enforced \cite{Rauss,qpl}, except that the effective holes are now carried by each $(\phi, \overline \phi)$ pair. Our scheme uses these to store quantum information.

States with $(\phi, \overline \phi)$ or $\lambda$ pairs on vertices connected by a single edge, $i$, can be created by acting on the ground state with the operators:
\bq \label{creation}
W^{\phi}_{i} = \frac{1}{2} \sigma_i^z \left[ 1 + (\sigma_i^z)^3 \right], \,\, W^{\lambda}_{i} = (\sigma_i^z)^3,
\eq
respectively. The projection $(1 + (\sigma_i^z)^3)/2$ present in $W_i^\phi$ may be performed deterministically by measuring the observable $(\sigma_i^z)^3$ and applying $\big(A (v)\big)^3$ to a neighboring vertex if the $-1$ eigenvalue is obtained. By considering the action of the creation operators on two edges sharing a vertex, one obtains the fusion rules,
\bq \label{frules} \nonumber
\phi \times \overline \phi = 1+\lambda, \quad \phi \times \lambda &=& \phi, \quad \overline \phi \times \lambda = \overline \phi, \\
\overline \phi \times \overline \phi = \phi, \quad \phi \times \phi &=& \overline \phi, \quad \lambda \times \lambda = 1.
\eq
In other words, if both a $\phi$ and $\overline \phi$ are moved into the same vertex $v$, their state can be stabilized by either $P_{1}(v)$ or $P_{\lambda}(v)$. Note that $W^{\phi}_{i} W^{\lambda}_{i} = W^{\phi}_{i}$, ensuring that the fusion of a $\lambda$ with a $\phi$ cannot be locally distinguished from a single $\phi$. This provides a non-trivial fusion space where information can be encoded in a delocalized way. The $\lambda$ particles can be transported using chains of $W^{\lambda}$, whereas $\phi$ and $\overline \phi$ require controlled operations \cite{Aguado,Wootton} such as,
\bq \label{C}
C_i = \sum_{g,h \in Z_6} P_{e_{g+h}} (v) P_{e_h} (v') \, (\sigma_i^z)^g,
\eq
which moves the particle coherently from vertex $v$ to the neighboring vertex $v'$ through the edge $i$. Alternatively, our quasiparticles are well suited to be moved adiabatically using local potentials \cite{me}, without affecting the degeneracy of the logical states.

Quasiparticles $\chi$, $\overline \chi$ and $\mu$ on plaquettes can be defined equivalently to $\phi$, $\overline \phi$ and $\lambda$, respectively. The corresponding projectors $P_\chi(p)$, $P_{\overline \chi}(p)$ and $P_\mu(p)$ and the creation operators $W_i^\chi$ and $W^\mu_i$ are obtained from \rf{P}, \rf{creation} and \rf{C} using the substitutions $A (v) \to B (p)$ and $\sigma_i^z \to \sigma_i^x$. The braiding of the quasiparticles can be determined from the constituent $e_g$ and $m_g$ anyons. For example, a $\mu$ around a $\lambda$ gives the statistical phase $e^{i\pi}$ due to their identification with $m_3$ and $e_3$, respectively.

\begin{figure}[t]
\begin{center}
{\includegraphics[scale=.55]{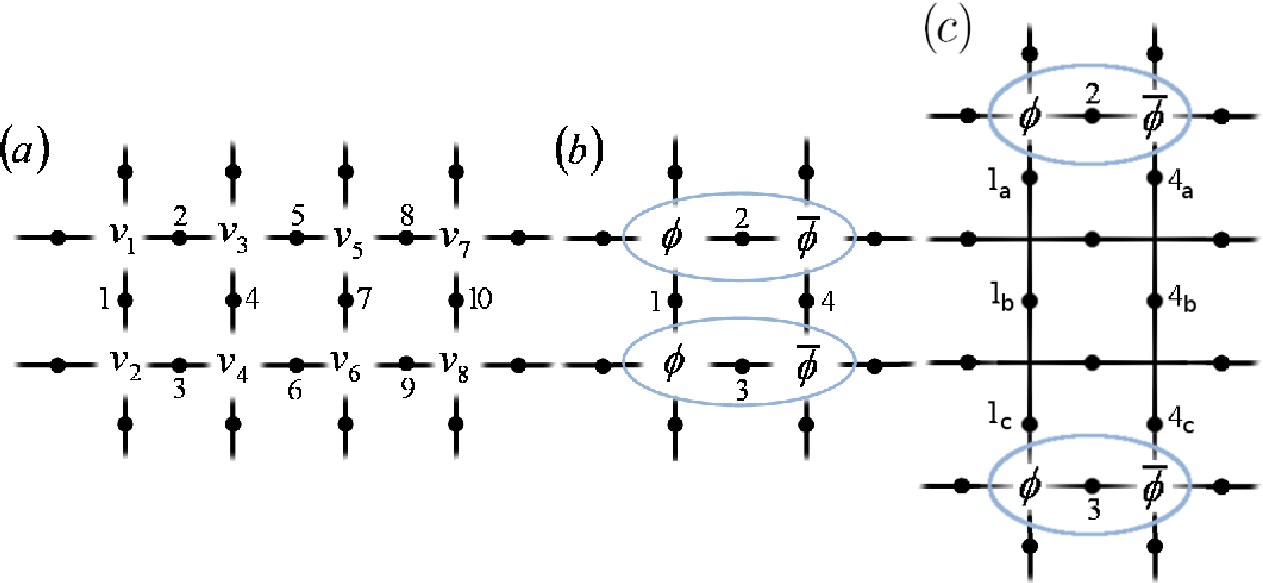} }
\caption{\label{figure1} (a) Part of the lattice with edges and vertices
enumerated. (b) The $(\phi, \overline \phi)$ pairs of quasiparticles in which a logical qubit may
be stored, with the pairs highlighted. The $Z$ basis may be measured by determining the fusion outcome,
$1$ or $\lambda$ of pairs connected by horizontal edges. The $X$ basis may
be accessed by single spin operations on vertical links. (c) The same scenario as (b), but with the two pairs separated. $X$ basis operations now have to act on more than one spin, such as three in this case.}
\end{center}
\end{figure}

\section{Quantum computation}

Employing the new quasiparticle states, we define a delocalized encoding equivalent to quantum computational schemes with non-Abelian anyons \cite{Anyons1,Double1,tqc,thesis}. Consider the following operations on the ground state (see Fig.\ref{figure1}(a) for enumeration): (i) The application of $W_2^\phi W_3^\phi$; (ii) The application of $W_1^\lambda$. Applying operation (i) creates two $(\phi, \overline \phi)$ pairs at vertices $v_2$, $v_4$, and $v_1$, $v_3$. The fusion of either pair would result in the vacuum. Applying both (i) and (ii) also creates two $(\phi, \overline \phi)$ pairs, but in this case they would both fuse to a $\lambda$. These states can be used to encode a $v$-type logical qubit, with basis states $\ket{0_v}$ and $\ket{1_v}$ associated with the vacuum and $\lambda$ fusion channels, respectively. In terms of spin operators acting on the ground state they can be written as,
\bq \label{} \label{fchannels}
 \ket{0_v} & = & \frac{1}{4} \sigma^z_2 \sigma^z_3 \left[1 + (\sigma^z_2)^3 \right] \left[ 1 + (\sigma^z_3)^3 \right] \ket{\textrm{gs}}, \nonumber \\
 \ket{1_v} & = & \frac{1}{4} (\sigma^z_1)^3 \sigma^z_2 \sigma^z_3 \left[ 1 + (\sigma^z_2)^3 \right] \left[1 + (\sigma^z_3)^3 \right] \ket{\textrm{gs}}.
\eq
We may also write these in terms of the anyon states at each vertex as follows,
\bq \label{} \label{fchannels2}
 \ket{0_v} & = & \frac{1}{4} (\ket{e_1}_{v_1} \ket{e_5}_{v_3} + \ket{e_4}_{v_1} \ket{e_2}_{v_3})(\ket{e_1}_{v_2} \ket{e_5}_{v_4} + \ket{e_4}_{v_2} \ket{e_2}_{v_4}), \nonumber \\
 \ket{1_v} & = & \frac{1}{4} (\ket{e_4}_{v_1} \ket{e_5}_{v_3} + \ket{e_1}_{v_1} \ket{e_2}_{v_3})(\ket{e_4}_{v_2} \ket{e_5}_{v_4} + \ket{e_1}_{v_2} \ket{e_2}_{v_4}).
\eq
It can then be easily verified using the rules in Eq. \rf{fuse} that fusing the $\phi$ (that is the $e_1$ or $e_4$) residing at $v_1$ with its corresponding $\overline\phi$ (the $e_2$ or $e_5$) at $v_3$ will always result in the vacuum for the state $\ket{0_v}$ and a $\lambda$ (or $e_3$) for $\ket{1_v}$. The same is true for the $\phi$ and $\overline \phi$ at $v_2$ and $v_4$.

No local operator (4-local such as $A(v)$ or smaller) can distinguish between the two states. The fusion channel of a $(\phi, \overline \phi)$ pair is a delocalized property, and can only be detected by observables acting on both particles of a pair. The logical qubit operations are given by,
\bq \nonumber \label{XZ}
X  &=&  (\sigma^z_1)^3 \, \textrm{or} \, (\sigma^z_4)^3, \\
Z &=& A^3 (v_1) A^3 (v_3) \ \textrm{or} \ A^3 (v_2) A^3 (v_4).
\eq
If the quasiparticles are moved the logical operations must change accordingly. Consider keeping the $\phi$ and $\overline \phi$ of each pair together, but using \rf{C} to move the two pairs away from each other. The $X$ operations become products of $(\sigma^z_1)^3$'s along strings connecting the targeted vertices, while the $Z$ operation remains the same. For example, consider the case in Fig.\ref{figure1}(c). Any operator on the single spin $1$ becomes an operator on the three spins $1_a$, $1_b$ and $1_c$, which lie on the corresponding path between the two $\phi$'s. Accordingly, the $X$ operations become,
\bq \nonumber \label{XZ}
X  &=&  (\sigma^z_{1_a})^3(\sigma^z_{1_b})^3(\sigma^z_{1_c})^3 \, \textrm{or} \, (\sigma^z_{4_a})^3(\sigma^z_{4_b})^3(\sigma^z_{4_c})^3.
\eq
Further separation will increase the lengths of the paths between pairs, and hence increase number of spins that need to be acted upon. This gives a topological protection against $X$ errors for large distances.

The two possible forms of both $X$ and $Z$ allow two possible means to measure in each basis. These should both give the same result, with any difference being a signature of errors. The fusion channels of $\chi$'s and $\overline \chi$'s may similarly be used to encode $p$-type qubits on corresponding plaquettes $p_1, \ldots, p_4$, with logical states $\ket{0_p}$ and $\ket{1_p}$. The $X$ and $Z$ rotations for the $p$-qubits are obtained using the substitutions above. The eigenstates of the $X$ basis may be denoted as $\ket{\pm_{v/p}}$.

\begin{figure}[ht]
\begin{center}
{\includegraphics[scale=.18]{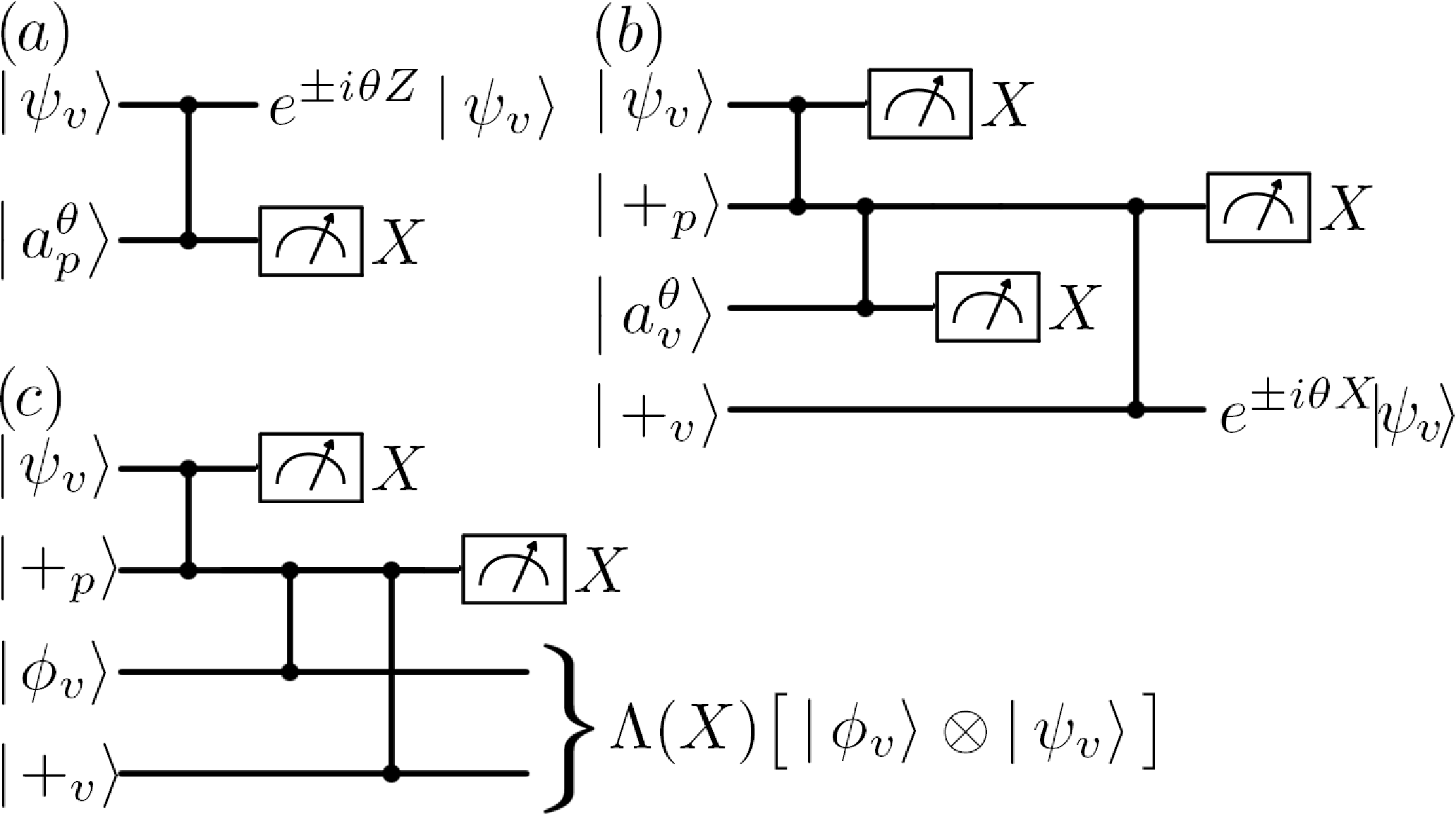} }
\caption{\label{figure2}The circuits that implement (a) $e^{\pm i \theta Z}$ rotations and
(b) $e^{\pm i \theta X}$ rotations. The sign depends on the outcome of the
measurements and can be corrected by subsequent rotations. In (c) the
controlled-$X$ gates are depicted on $v$-type qubits. The $X$ measurements
here may be realized by single spin measurements $(\sigma^z_i)^3$ for
$v$-type qubits and $(\sigma^x_i)^3$ for $p$-type.}
\end{center}
\end{figure}

The controlled-$Z$ gate can be applied between a $v$-type and a $p$-type qubit by moving the $(\phi, \overline \phi)$ pair on $v_1$, $v_3$ around the $(\chi, \overline \chi)$ pair on $p_1$, $p_3$. This is due to the statistical phase $e^{i\pi}$ obtained when moving a $\lambda$ around a $\mu$. Arbitrary single qubits rotations can be performed by employing suitable logical ancillary states. To do this we initially prepare the logical state $\ket{0_v}$. The projector $\Pi^{\theta}_1 = \frac{1}{2} \left[ 1 + \cos \theta \, (\sigma^x_1)^3+ i \sin \theta \, (\sigma^z_1)^3(\sigma^x_1)^3 \right]$ is then applied to the spin on edge $1$. This can be performed probabilistically, by a measurement. In general, $\Pi^{\theta}_1$ creates a superposition of a pair of $\lambda$'s and $\mu$'s on the vertices and plaquettes sharing the spin on edge $1$. The $\lambda$'s will fuse with the $\phi$'s, and the $\mu$'s may be measured. If a pair of $\mu$'s is detected, they can be annihilated by applying $(\sigma^x_i)^3$. This leaves the system in the logical ancillary state $\ket{a^\theta_v} = (\cos \theta \ket{0_v} - i\sin \theta \ket{1_v})/{\sqrt{2}}$. If no $\mu$'s are detected, the protocol can be repeated until successful. The ancillary plaquette state $\ket{a^\theta_p}$ can be prepared similarly. By utilizing these states, arbitrary single qubit rotations and controlled-$X$ gates can be performed using the circuits shown in Fig. (\ref{figure2}).

During the preparation stage, the ancilla states are stored on neighboring plaquettes or vertices, making them vulnerable to $X$ errors and affecting the fault-tolerance of our scheme. However, note that all single qubit rotations can be constructed from $e^{i (\pi / 8) Z}$ and $e^{i (\pi / 8) X}$ \cite{Uni}, whose implementation requires only preparation of $\ket{a^{\pi /8}_v}$, $\ket{a^{\pi /8}_p}$ and $\ket{+_{p/v}}$. The former two may be prepared fault-tolerantly from many noisy copies by use of distillation \cite{Magic,Rauss}. The latter is an eigenstate of $X$, and so intrinsically resilient to these errors. The controlled-$X$ also requires only these ancilla states, completing the fault-tolerant universal gate set.  An alternative gate set can be performed using non-topological operations on the underlying spins \cite{tqc}.

\section{Fault-tolerance}

We consider errors that do not excite the system, requiring a temperature low enough for topological order to be stable \cite{Castelnovo,Iblisdir}. The errors can then be considered as perturbations in the Hamiltonian, due to imprecise tuning of the system or coupling with the environment.

The Hamiltonian \rf{H} can be expressed as an equally weighted sum of the stabilizers $A(v)$ and $B(p)$. However, physical systems will likely produce perturbed Hamiltonians, lifting the degeneracy of the anyons and breaking the symmetries our scheme requires. This is a problem that not only affects realizations of $D(Z_6)$, but all quantum double models, Abelian or non-Abelian \cite{Double1,Aguado,tqc}. Here we present a method to enforce the symmetries in $D(Z_6)$, but the principle applies in general. 

We require that the states within the computational space, as stated in Eq. \rf{fchannels2} for logical qubits stored on vertices, are unaffected by perturbations that lift the degeneracies of the $e_1$ and $e_4$ anyon states and the $e_2$ and $e_5$ states. If this is not so, relative phases will accumulate between the superposed anyon states, causing errors in $X$ basis measurements. For example the perturbation $\delta A^3(v_1)$ on vertex $v_1$ does not commute with $W^{\lambda}_1$. Hence measuring the logical $X$ will give the wrong result.

To see how these perturbations may be dealt with, consider spins $2$ and $3$. According to the definition of the logical states in Eq. \ref{fchannels}, the projection $(1+W^{\lambda}_i)$ is applied to each of these spins. This makes the logical states eigenstates of $W^{\lambda}_2$ and $W^{\lambda}_3$, both with eigenvalue $+1$. The addition of the single spin terms $-B W^{\lambda}_2$ and $-B W^{\lambda}_2$ to the Hamiltonian therefore has no effect on states within the computational space except to reduce their energy. However, since the terms anticommute with any perturbations that lift the required degeneracies, these are energetically suppressed by a gap $2B$. For example, a perturbation of strength $\delta$, such as $\delta A^g(v)$,  is suppressed by $(\delta/B)^2$. The polynomial suppression of errors is not as efficient as would usually be expected in a topological model. However, it is reasonable to expect that the single spin $B$ can be made to be much greater than the many-body perturbations $\delta$. Errors will then be greatly suppressed. An equivalent method for the $(\chi, \overline\chi)$ pairs can be obtained with $-B W^{\mu}_i$ terms.

Moving the quasiparticles of a pair apart means that the magnetic field term is no longer a single $(\sigma^z_i)^3$ but a product along a path stretching between them. This will make it harder to implement, reducing the strength that may be achieved and so reducing its effectiveness. As an alternative, each pair can be replaced by a bank of $N$ pairs in a similar way to that described in \cite{thesis,qpl}. This then gives $2N$ independent forms for the logical $X$ operator, and so $2N$ independent measurements that may be made. Using majority voting on these when measuring the $X$ means that the probability of an error will be exponentially suppressed with $N$.

We now consider perturbations that do not come from fine tuning, such as those acting on spins forming strings across the lattice. If one end of a string connects with a $\phi$ or $\overline \phi$, it may move a quasiparticle or cause it to annihilate. The simplest examples are single spin perturbations $(\sigma^z_i)^g$, acting on a spin surrounding a $\phi$ or $\overline \phi$. These can be suppressed by changing the the Hamiltonian terms on vertices in which a $\phi$ or $\overline \phi$ resides to $P_\phi (v)$ and $P_{\overline\phi} (v)$, respectively. This energetically favours the quasiparticles being located at these points, and so stops them moving or annihilating in error.

If string-like perturbation stretches between the two $(\phi, \overline \phi)$ pairs, it can distinguish the logical states of the $X$ basis and so lift the degeneracy. If the environment can produce string-like $k$-local perturbations in the Hamiltonian then the $(\phi, \overline \phi)$ pairs should be moved $k+1$ spins apart, to protect the encoded qubit. Perturbations may also act on spins that form loops. If these loop around a single $\phi$ or $\overline \phi$, they may cause errors on the results of $X$ measurements. These will be suppressed by the magnetic field $(\sigma^z_i)^3$ as long as they do not loop around both the $\phi$ and $\overline \phi$ of a pair. Such errors become highly correlated, and so increasingly difficult for nature to produce, as these particles are moved apart, or if each pair is replaced by banks of $N$ pairs \cite{thesis,qpl}.

\section{Josephson junction realization}

\begin{figure}[t]
\begin{center}
{\includegraphics[scale=.23]{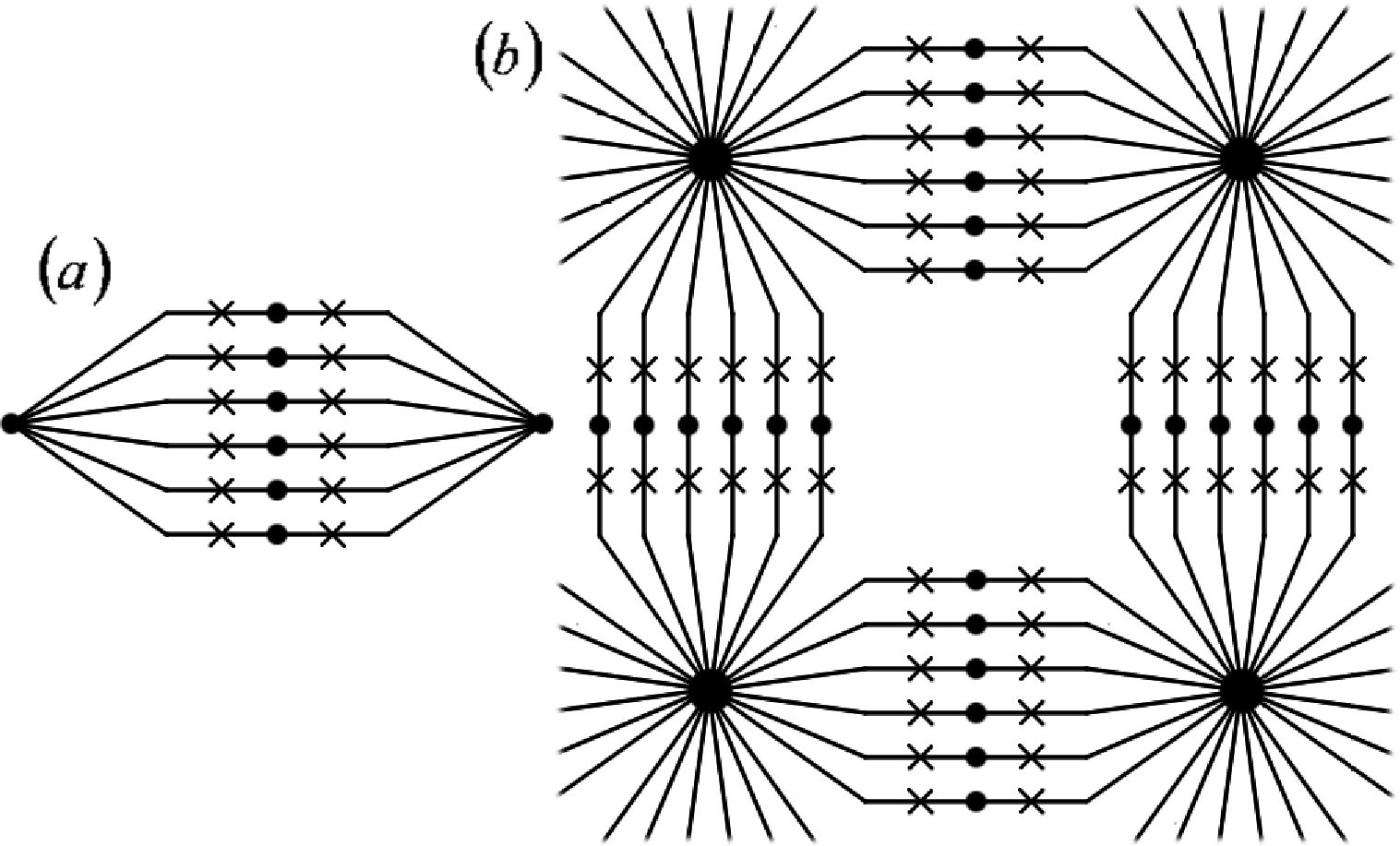} }
\caption{\label{figure3} (a) Twelve Josephson junctions arranged in five loops. This element has six degenerate ground states when a flux $2 \pi /6$ is passed through each loop, and thus provides the required six-level spin. (b) A plaquette of the $D(Z_6)$ model, equivalent to that shown in Fig.\ref{figure1}(b), realized with Josephson junction elements.}
\end{center}
\end{figure}

We now present a means to experimentally demonstrate the quantum memory of our scheme, including its resilience to perturbations breaking the required symmetries. Consider the Josephson junction element in Fig. \ref{figure3}(a). A flux $2 \pi /6$ passes through each loop, creating six degenerate ground states that can be used as a six-level spin \cite{Doucot}. Constructing a lattice of such elements, as in Fig. \ref{figure3}(b), imposes the vacuum state on all plaquettes, and gives the following Hamiltonian for the vertices,
\be
H'=- r \sum_v \big(A (v) + A^\dagger (v)\big).
\ee
Here $2r$ is the energy gap resulting from the tunnelling processes within the junctions. Using a semi-classical approximation we find the coupling to be $r \approx E_J^{3/4} E_C^{1/4} \exp(-S_0)$ with $S_0 \approx 0.380 \sqrt{E_J/E_C}$. Here $E_J$ and $E_C$ are the Josephson and charging energies, respectively. To realize a single qubit memory, the states \rf{fchannels} must be prepared. This can be done by pumping charge between vertices to implement the $\sigma^z_i$ operations on the connecting link \cite{Pump}. The degeneracy of these states must then be confirmed. Since the Hamiltonian is a perturbed version of \rf{H}, it does not assign equal energies to the $e_g$ anyons. Hence, the terms $B (\sigma^z_2)^3$ and $B (\sigma^z_3)^3$ must be applied to each pair to ensure the degeneracies. These can be simply implemented by passing a flux of $2 \pi /3$ through the elements on links $2$ and $3$, rather than $2 \pi /6$. 

The models used for our scheme are also closely related to those demonstrated with optical experiments \cite{Pachos,Pan}, giving another possible avenue for experimental study. The experimental verification of the memory of our scheme by any means, even without the application of any quantum gates, would form a major breakthrough in the realization of anyonic quantum memories.

\section{Conclusions}

We have demonstrated non-Abelian like encoding of quantum information using the more experimentally accessible Abelian anyon models. Since the scheme requires a fine tuned Hamiltonian, we have also introduced a method to enforce symmetries that can be applied to other models, both Abelian and non-Abelian. Further, we have proposed an experiment to demonstrate the quantum memory and enforcing of symmetries with cutting edge technology. It would be interesting to study whether the enhanced fault-tolerance provided matches that of non-Abelian schemes requiring non-topological operations for universality \cite{Magic}.


\section{Acknowledgements}

We would like to thank Gavin Brennen and David DiVincenzo for inspiring discussions and for critical reading of the manuscript. This work was supported by the EU grants EMALI and SCALA, the EPSRC, the Finnish Academy of Science and the Royal Society.

\end{document}